\documentclass[aps,superscriptaddress,prb,groupaddress,twocolumn,floatfix]{revtex4-2}
\usepackage{units}
\usepackage{amsmath}
\usepackage{amssymb}
\usepackage{graphicx}
\usepackage{bm}
\usepackage{multirow, color,relsize,ulem,microtype}

\newcommand{\be}{\begin{equation}}
\newcommand{\ee}{\end{equation}}

\newcommand{\im}[1]{\text{Im}[#1]}

\begin{document}

\title{Forbidden $T$-symmetry breaking in time-dependent non-Hermitian systems \\with gain and loss} 

\author{Li Ge}
 \email{li.ge@csi.cuny.edu}
\affiliation{Department of Physics and Astronomy, College of Staten Island, CUNY, Staten Island, NY, USA}
\affiliation{The Graduate Center, CUNY, New York, NY, USA}

\begin{abstract}
Parity-time ($PT$) symmetry has attracted significant interest in photonics and related fields in recent years. With sufficiently weak gain and loss, these systems typically exhibit entirely real energy levels, with all corresponding wave functions remaining $PT$ symmetric. As the strength of gain and loss increases, complex-conjugate energy levels are expected to emerge, signaling the onset of $PT$ symmetry breaking. Similar properties can be found in non-Hermitian systems with other antilinear symmetries as well, where parity is replaced by another operator such as rotation ($R$) or simply the identity operator, resulting in $RT$-symmetric and $T$-symmetric non-Hermitian systems. In this work, we demonstrate that this typical scenario of non-Hermitian symmetry breaking in a finite Hilbert space is forbidden in a class of time-dependent systems with gain and loss and $T$ symmetry: the system remains in its $T$-symmetric phase as we increase the strength of gain and loss, even when the spatial distribution of gain and loss is random. Furthermore, this behavior is also  independent of the duration or period of the temporal modulation. In contrast, this forbidden non-Hermitian symmetry breaking does not occur when the same systems are modified to display $PT$ symmetry instead. Our findings highlight contrasting behaviors in time-dependent and time-independent non-Hermitian systems, creating new possibilities and opportunities for future investigations.
\end{abstract}

\maketitle

\section{Introduction}

Non‑Hermitian physics has emerged as a fertile arena for exploring optical systems in which gain and loss play an essential dynamical role \cite{bender_pt_2024}. A particularly influential class of such systems obeys parity–time ($PT$) symmetry, originally introduced in quantum mechanics as a generalization of Hermitian Hamiltonians with real energy levels. In a Schr\"odinger-like system such as paraxial photonics \cite{Feng2017Non-HermitianSymmetry}, $PT$ symmetry is realized by engineering the potential landscapes whose real part is even under a parity operation while their imaginary part is odd. When this condition is satisfied, the discrete energy levels can remain entirely real \cite{Bender1998}, enabling optical structures with balanced gain and loss to exhibit dynamics traditionally associated with conservative or unitary systems. 

Besides real energy levels, a more generic $PT$-symmetric Hamiltonian also displays pairs of complex conjugate energy levels, signaling that the corresponding wave functions are no longer eigenstates of $PT$. This transition between $PT$‑symmetric and $PT$-broken phases has been observed on experimental platforms including coupled waveguides, micro‑ring resonators, plasmonic systems, and metasurfaces \cite{miri_exceptional_2019}. It is marked by the emergence of exceptional points (EPs), which are non‑Hermitian degeneracies at which both eigenvalues and eigenvectors coalesce. These singularities give rise to a host of unconventional optical phenomena, including loss‑induced transparency, unidirectional reflectionless transport, non‑reciprocal mode conversion, and enhanced responses. 

With other system parameters fixed, this transition or non-Hermitian symmetry breaking in a finite Hilbert space (e.g., for a $PT$ dimer or chain \cite{ge_effective_2024}) typically takes place when the gain-loss strength is above a critical threshold, and this behavior is observed in non-Hermitian systems with other antilinear symmetries as well, i.e., where the symmetry operator incorporates the complex conjugation just as $PT$ does. Examples include $RT$-symmetric \cite{ge_parity-time_2014,ge_scattering_2015} and $T$-symmetric systems \cite{FNHSE}, where parity is replaced by rotation ($R$) and simply the identity operator, respectively. Such a non-Hermitian symmetry breaking is not limited to time-independent systems either, as shown by temporally modulated quantum walks \cite{xiao_observation_2017} and dissipative Floquet systems of ultracold atoms \cite{li_observation_2019}, where the effective Hamiltonian and its quasi-energies are defined via the evolution matrix for a given temporal period. 

In this work, we demonstrate that this typical scenario of non-Hermitian symmetry breaking is forbidden in a class of time-dependent systems with gain and loss and $T$ symmetry: the system remains in the T-symmetric phase, with an entirely real quasi-energy spectrum, as the strength of gain and loss increases, even when the spatial distribution of gain and loss is random. Furthermore, this behavior is also independent of the duration or period of the temporal modulation. In contrast, this forbidden non-Hermitian symmetry breaking does not occur when the same systems are modified to display $PT$ symmetry.

Below we first outline $T$ symmetry and illustrate its breaking in time-dependent non-Hermitian systems consisting of two non-Hermitian steps separated by three Hermitian ones. We then elucidate two scenarios of  forbidden $T$-symmetry breaking, and we attribute them both to a non-unitary transformation of the evolution matrix. In the first scenario, the gain and loss are correlated in each non-Hermitian step, and we allow non-staggered couplings in the Hermitian step between them. In the second scenario, the gain and loss in one non-Hermitian step can be completely random, but the couplings in the Hermitian steps must be staggered. Finally, we show that neither scenario can arise when we modify our systems to display $PT$ symmetry instead, i.e., with the two non-Hermitian steps mapped to each other by $PT$.

\section{Results}

\subsection{Time-dependent $T$ symmetry}

Our temporally modulated lattice system is defined by a time-dependent Hamiltonian $H(t)$ with $T$ symmetry. For simplicity, we consider classical and integer-spin systems where the $T$ symmetry takes the form 
\be 
TH(t)= H^*(t)T = H(-t)T.\label{eq:T}
\ee
Here, the asterisk indicates complex conjugation ($K$) as usual. If $H(t)$ is time-independent, then this condition leads to a real-valued Hamiltonian, which is also Hermitian unless the couplings are non-reciprocal, i.e., $H$ differs from its matrix transpose $H^T$. A time-dependent $H(t)$ allows complex matrix elements at any given time, and those appearing on the diagonal of $H(t)$ represent optical gain and loss in photonic systems, similar to the time-independent case \cite{Feng2017Non-HermitianSymmetry}.

Although we do not require the time-dependence to be periodic, we define the quasi-energy in the same way as in a Floquet system: by considering a time duration $D$, the quasi-energy $E$ is given by an eigenvalue of the effective Hamiltonian $H_{d}$ (with $\hbar=1$):
\be
U(D) = \mathcal{T}e^{-i\int_{-D/2}^{D/2} H(t) dt}  \equiv e^{-iH_{d}D}.  \label{eq:Hd}
\ee 
Here \( \mathcal{T} \) is the time-ordering operator, and $U$ is the temporal evolution operator. Below we focus on the imaginary parts of all quasi-energies, which indicate whether the eigenstates of $U(D)$ (and $H_d$) are amplified ($\im{E}>0$), attenuated ($\im{E}<0$), or stationary ($\im{E}=0$) as a function of time.

Below we drop the explicit $D$-dependent of $U$. By taking the complex conjugate of $U$, we find
\be
U^*U =  \mathcal{T} e^{i\int_{-D/2}^{D/2} H(-t) dt} e^{-i\int_{-D/2}^{D/2} H(t) dt} = 1 = UU^*. \label{eq:Tsymm}
\ee
The cancellation of the two exponential functions can be understood by dividing the duration $D$ into $2N$ equal intervals and letting $N\rightarrow\infty$. In each slice $H(t)$ can then be taken as time-independent, and we write them as $H_{-N},H_{-N+1},\ldots,H_{N-1}, H_N$. $U^*U$ is then given by 
\be
\lim_{N\rightarrow \infty} \left(e^{\frac{iH_{-N}D}{2N}} \ldots e^{\frac{iH_{N}D}{2N}} \right)
\left(e^{-\frac{iH_{N}D}{2N}} \ldots e^{-\frac{iH_{-N}D}{2N}} \right).
\ee 
The rightmost term in $U^*$ (i.e., $e^{\frac{iH_{N}D}{2N}}$) then cancels the leftmost term in $U$ (i.e., $e^{\frac{-iH_{N}D}{2N}}$), and the pairwise cancellation repeats for the remaining parts of this product. 

\subsection{$T$-symmetry breaking}

\begin{figure}[b]
\centering
\includegraphics[width=0.7\linewidth]{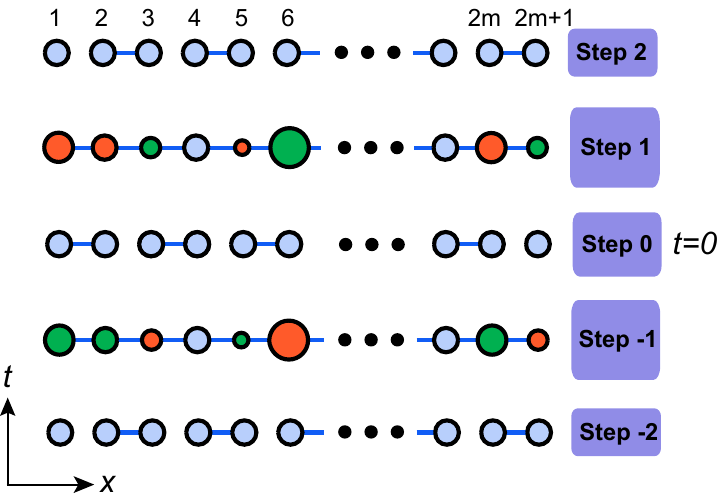}
\caption{\textbf{Example of a time-dependent non-Hermitian chain with $T$ symmetry}. Lines indicate real and reciprocal couplings, and green (red) dots of different sizes represent gain (loss) of various strength in steps $-1$ and 1. The numbers on top mark the positions of the sites, with $m\in\mathbb{Z}$.}
\label{Fig:1}
\end{figure}

Equation~(\ref{eq:Tsymm}) indicates that if $\psi_\mu$ is an eigenstate of $U$, i.e., $U\psi_\mu = \lambda_\mu\psi_\mu$, then we find $U^*U\psi_\mu = \lambda_\mu U^*\psi_\mu = \psi_\mu$, or equivalently,
\be
 U\psi^*_\mu = \frac{1}{\lambda^*_\mu}\psi^*_\mu.
\ee 
This relation indicates that $\psi^*_\mu$ is also an eigenstate of $U$, and there are two possible ``phases'' of $\psi_\mu$. In the $T$-symmetric phase, we have
\be 
T\psi_\mu = \psi^*_\mu = \psi_\mu, \quad |\lambda_\mu|=1, 
\ee 
i.e., the eigenvalues of $U$ are on the unit circle in the complex plane. In the $T$-broken phase, we have instead 
\be 
T\psi_\mu = \psi_\nu, \quad \lambda_\mu^*=\lambda_\nu^{-1},
\ee 
and $\lambda_\mu,\lambda_\nu\in\mathbb{C}\,(\mu\neq\nu)$ have the same phase angle and reciprocal amplitudes. These properties are similar to a time-independent scattering matrix with $PT$ and $RT$ symmetries \cite{chong_pt-symmetry_2011,ge_scattering_2015}. With $T$ symmetry (i.e., without gain and loss), however, a time-independent scattering matrix is unitary and always has $|\lambda_\mu|=1$ due to energy conservation.

Next, we note that $U$ and the effective Hamiltonian $H_d$ share the same eigenstates [see Eq.~(\ref{eq:Hd})], and their eigenvalues satisfy $\lambda_\mu = e^{-iE_\mu D}$. Therefore, the corresponding $T$-symmetric and $T$-broken phases of $H_d$ are both characterized by 
\be 
E_\mu=E^*_\nu, 
\ee 
and we find $\mu=\nu$ ($\mu\neq\nu$) in the $T$-symmetric ($T$-broken) phase. This is similar to a $PT$-symmetric Hamiltonian, whether it is time-independent \cite{Feng2017Non-HermitianSymmetry} or not \cite{xiao_observation_2017,li_observation_2019}.

To exemplify these two phases, we utilize the 1D chain shown in Fig.~\ref{Fig:1}, which is motivated by studies in Floquet topological phases \cite{rudner_anomalous_2013} and non-Hermitian skin effect \cite{FNHSE}. It has an odd number of sites and consists of five steps. The Hamiltonian is time-independent in each step, and the durations of these steps satisfy $D_{-n}=D_{n}\,(n=1,2)$, with $t=0$ at the middle of step 0 with duration $D_0$.  

Steps $-2$ and $2$ are the same and feature real and staggered nearest-neighbor (NN) coupling $C$, i.e., between sites 2 and 3, 4 and 5, and so on. Step $0$ has the same real NN coupling $C$ but staggered differently, i.e., between sites 1 and 2, 3 and 4, etc. Note that this real $H_0$ is its own time reversal, i.e., $H_0=H_0^*$. 

Step $-1$ has random real NN couplings everywhere and a random distribution of gain and loss, and step 1 is its time reversal with gain (loss) replaced by loss (gain) of the same amount, represented by an imaginary on-site potential with opposite signs. 

\begin{figure}[b]
\centering
\includegraphics[width=\linewidth]{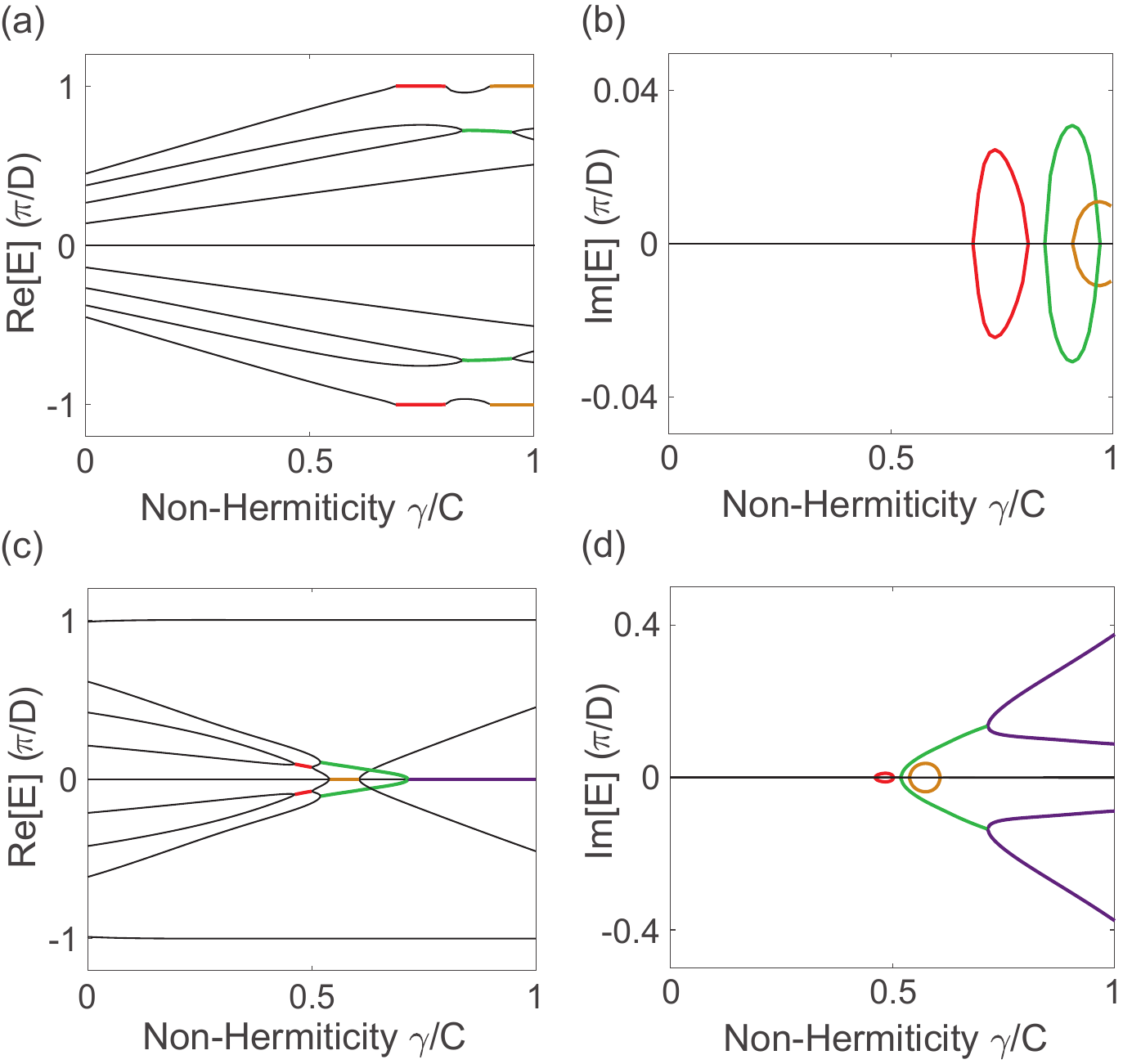}
\caption{\textbf{$T$-symmetry breaking in a time-dependent non-Hermitian chain}. (a,b) Real and imaginary parts of quasi-energies in the system shown in Fig.~\ref{Fig:1} with 9 sites. Here $D=3/C$ and $D_{i}=D/8\,(i=-2,2),\,D/4\,(i=-1,0,1)$. $\gamma$ is the maximal strength of gain and loss, and we increase it by scaling $H_{-1}$ and $H_1$ as a whole. (c,d) Same as (a,b) but with $D=9/C$ and another random $H_{-2}$ (and $H_{2}$).
}
\label{Fig:2}
\end{figure}

To enhance $T$-symmetry breaking, below we scale $H_{-1}$ and $H_1$ as a whole when increasing the non-Hermiticity, instead of just scaling the gain and loss they include. We use the maximal strength of gain and loss to represent the non-Hermiticity, and we denote it by $\gamma$. 

In the limit that $D\rightarrow0$, the temporal evolution operator $U$ approaches an identity matrix, and hence all quasi-energies converge to 0 on the zeroth Riemann surface of the logarithm. With a finite but small $D$, the real quasi-energies still cluster around 0 when $\gamma=0$ [see Fig.~\ref{Fig:2}(a)]. As we increase $\gamma$, we observe that the quasi-energies fan out and stay real initially, conforming with the typical scenario of non-Hermitian symmetry breaking mentioned in the introduction. 

In addition, they seem to satisfy $E_\mu=-E_{\mu'}$, which is not warranted by $T$ symmetry of the system. We attribute this property to $E_\mu=-E^*_{\mu'}$ instead, a result of non-Hermitian particle-hole (NHPH) symmetry \cite{Ge2017,qi_defect_2018,schomerus_topologically_2013,kawabata2019topological,okugawa2019topological}. It is defined by
\be 
\{H_d,CK\} = 0, \label{eq:NHPH1}
\ee  
where $K$ is again complex conjugation and $C$ is the diagonal sublattice operator with alternate 1's and $-1$'s in position space. In the simplest case, NHPH symmetry holds for a time-independent system with two sublattices and random gain and loss, where couplings are real and exist only between two sites on different sublattices.

These conditions are satisfied for each of the five time-independent Hamiltonians $H_i$ $(i=-2,-1,0,1,2)$ in this case, and hence we have $\{H_i,CK\}=0$, and in turn, $CK H_i^n = (-1)^n H_i^n CK\,(n\in\mathbb{Z})$. The corresponding evolution matrix then satisfies $[U_i,CK]=0$. 
Note that the switch from an anticommutation relation to a commutation relation is due to the imaginary unit ``$i$'' in the exponent of $U_i$. Therefore, we also find 
\be 
[U,CK]=0, \label{eq:NHPH2}
\ee     
which then gives the NHPH symmetry of $H_d$ shown by Eq.~(\ref{eq:NHPH1}) using $C^2=1$ \cite{bibnote:1}.

Equation~(\ref{eq:NHPH2}) indicates that the eigenvalues $\lambda$'s of $U$ form complex conjugate pairs. Therefore, when two $\lambda$'s meet at $-1$ on the unit circle, they give rise to two $\pi$-modes of $H_d$ that undergo a $T$-breaking transition. This behavior takes place at $\gamma\approx0.69C$ in Figs.~\ref{Fig:2}(a,b), and if we further increase $\gamma$ (upto $0.80C$), the real parts of their quasi-energies remain at $\pm \pi/T$ in the $T$-broken phase while their imaginary parts become nonzero and opposite [red sections in Figs.~\ref{Fig:2}(a,b)]. The $T$ symmetry of these two modes can restore and break again once as we increase $\gamma$ even further, while the other modes also start to undergo $T$-symmetry breaking [green sections in Figs.~\ref{Fig:2}(a,b)]. 

If we start with a large $D$ where the Hermitian system already has two topologically protected $\pi$-modes \cite{nathan_topological_2015} (which, again, does not require $H(t)$ to be periodic in time), a typical scenario is that these two modes do not undergo a $T$-breaking transition as we increase $\gamma$ [Fig.~\ref{Fig:2}(c,d)]. In the meanwhile, the other modes converge toward $E=0$ before fanning out again, and they can experience multiple $T$-breaking transitions.

If the system only has a ``local'' instead of a ``global'' $T$ symmetry, e.g., with the system shown in Fig.~\ref{Fig:1} experiencing an additional unitary evolution either before or after the $T$-symmetric $H(t)$, then the two contrasting behaviors of the quasi-energies in the $T$-symmetric and $T$-breaking phases are lost: The effective Hamiltonian of the entire temporal evolution has complex quasi-energies in general as soon as $g$ is nonzero (not shown). Nevertheless, they may still have other symmetry properties, such as NHPH symmetry discussed above. \\

\subsection{Forbidden $T$-symmetry breaking}

If we implement gain and loss in the absence of couplings, i.e., removing the latter in steps $-1$ and $1$ of the system shown in Fig.~\ref{Fig:1}, we find a completely different behavior: the quasi-energies remain real even for large non-Hermiticity $\gamma$, and they are \textit{unchanged} from their Hermitian counterparts (with $\gamma=0$) and hence also independent of the random spatial distribution of gain and loss in steps $-1$ and $1$. 

\begin{figure}[t]
\centering
\includegraphics[width=\linewidth]{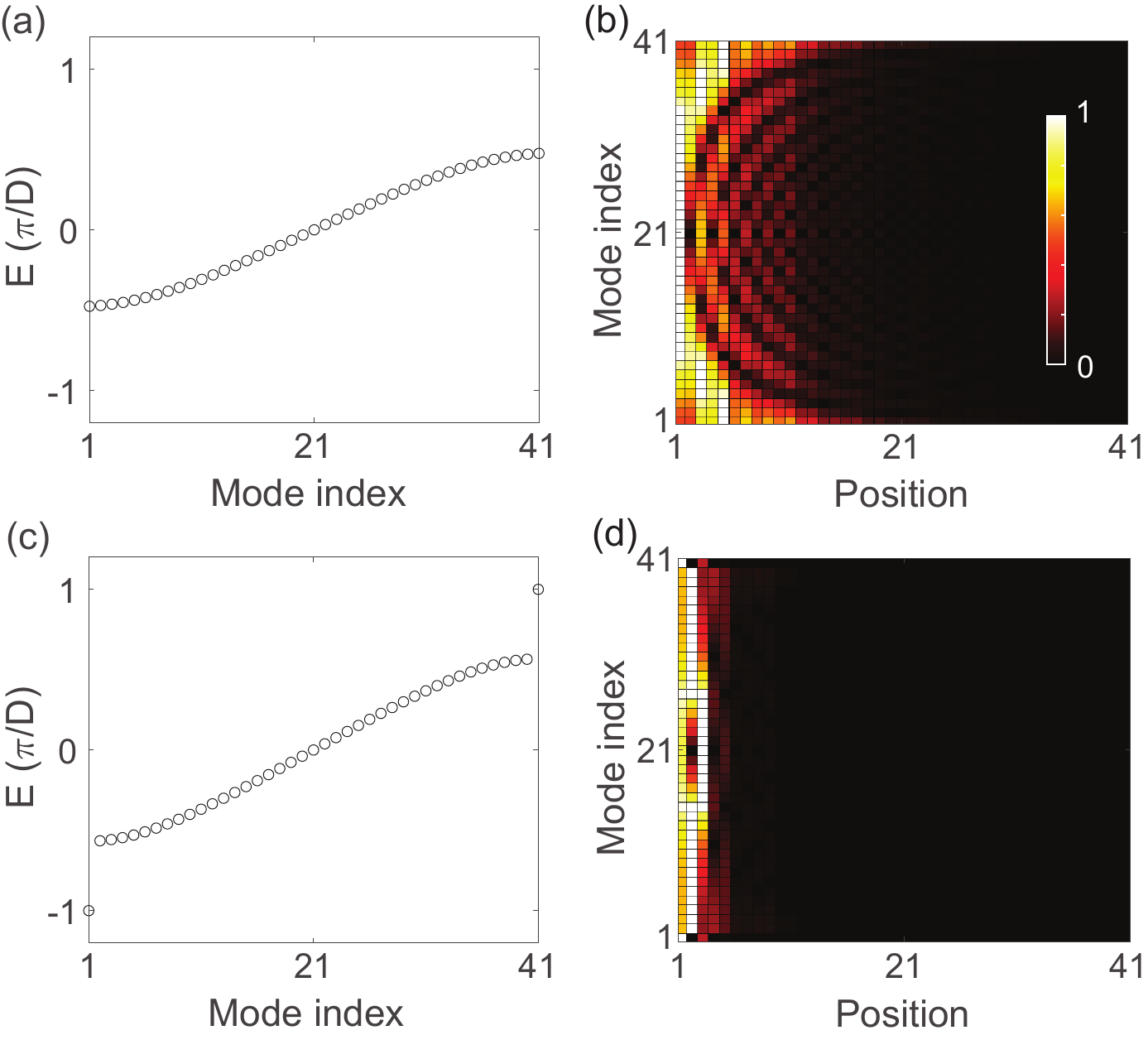}
\caption{\textbf{Forbidden $T$-symmetry breaking in a time-dependent non-Hermitian skin chain}. (a,c) Real quasi-energies of the system shown in Fig.~\ref{Fig:1} with 41 sites and no coupling in steps $-1$ and $1$. Here $\gamma=1$ and $D=3/C$ and $9/C$, respectively. (b,d) Spatial profile $|\psi(x)|$ of all eigenmodes of $H_d$ in (a,c).
}
\label{Fig:3}
\end{figure}

We exemplify this forbidden $T$-symmetry breaking in Figs.~\ref{Fig:3}(a,c), both with and without the $\pi$-modes. A similar configuration but with uniform gain and loss was used in Ref.~\cite{FNHSE} to study the non-Hermitian skin effect and its frustration with a Floquet drive protocol. Here we also observe the non-Hermitian skin effect across all modes [Figs.~\ref{Fig:3}(b,d)], with only gain on one sublattice in step $-1$ and only loss on the other sublattice in step 1.  

To gain analytical insights of $H_d$ and the forbidden $T$-symmetry breaking, one could use, for example, the Baker–Campbell–Hausdorff (BCH) formula 
\be 
e^{a}e^{b}\approx e^{a+b+[a,b]/2}\label{eq:BCH}
\ee 
to first find the effective Hamiltonians $H_-$ and $H_+$ of the two half periods, i.e., 
\begin{align} 
e^{-i H_{-} D/2} &\equiv U_0\left(\frac{D_0}{2}\right)U_{-1}(D_{-1})U_{-2}(D_{-2}),\\
e^{-i H_{+} D/2} &\equiv U_2(D_2)U_1(D_1)U_0\left(\frac{D_0}{2}\right), 
\end{align}
and then use this formula again to find $H_d$ from
\be
 e^{-i H_{d} D} = e^{-i H_{+} D/2}e^{-i H_{-} D/2}.
\ee

Via an imaginary gauge transformation between $H_d$ and a Hermitian Hamiltonian \cite{Hatano1996LocalizationMechanics}, this approach was able to show that all quasi-energies of $H_d$ are real with uniform gain and loss \cite{FNHSE}. However, it is only an approximate method that applies when $CD\ll1$ and $\gamma D\ll1$. Furthermore, these real quasi-energies are expected from the typical case of non-Hermitian symmetry breaking, i.e., where the spectrum remains real for weak non-Hermiticity. 

To prove that $T$-symmetry breaking is forbidden in this case with arbitrary strength and spatial distribution of gain and loss, an exact approach is required. If we inspect the matrix elements of $H_d$ for large $\gamma$, we find that the longer-range couplings become non-negligible, which can even be stronger than the NN couplings. While this observation makes the forbidden $T$-symmetry breaking even more striking, it also indicates that deriving an analytical form of $H_d$ directly will be challenging. Therefore, we turn out attention to $U$ instead. 

Without couplings in steps $-1$ and $1$, we note that $H_{-1}$ and $H_1$ are both diagonal, so are $U_{-1}$ and $U_1$. They satisfy $H_1=-H_{-1}$ due to the time reversal of gain and loss, and in turn, 
\be 
U_{1}U_{-1}=e^{-iH_1D/4}e^{iH_1D/4} = 1.
\ee 
Therefore, we find $U_{-1}=U_1^{-1}$, which allows us to rewrite $U$ as
\be
U = U_2(U_{1} U_0 U_{1}^{-1}) U_{-2}.\label{eq:U}
\ee 
In other words, $U_1^{-1}$ performs a similar transformation of $U_0$ in $U$. 
However, this similar transformation does \textit{not} guarantee that $U$ shares the same eigenvalues with the unitary evolution  
\be
U_q = U_2U_0U_{-2} \equiv e^{-iH_qD}, \label{eq:Hermitian}
\ee
where $H_q$ is Hermitian. 
Only if a similar transformation exists between $U$ itself and $U_q$ can we conclude that they have the same eigenvalues, i.e., on the unit circle in the complex plane. In turn, the quasi-energies of $H_d$ are then guaranteed to be real, which leads to forbidden $T$-symmetry breaking of our time-dependent non-Hermitian lattice.

We search for this similar transformation by inspecting the structure of each individual evolution matrix in all  steps, and we analyze them below in the sublattice basis, where sublattice $\mathbb{A}$ has $(m+1)$ sites and sublattice $\mathbb{B}$ has $m$ sites, with a total of $M=2m+1$ sites. 

First, we denote the diagonal of $U_1$ with random gain and loss by
\be 
[U_1]_i = r_i\,(i\in\mathbb{A}),\, g_i\,(i\in\mathbb{B}). \label{eq:U1}
\ee 
$r_i$ and $g_i$ represent the non-Hermiticity on sublattice $\mathbb{A}$ and $\mathbb{B}$ respectively, and they are less (greater) than 1 if there is loss (gain) at site $i$ in the corresponding sublattice. 

The staggered and reciprocal couplings in step 0 leads to 
\be 
U_0 = S_0 + 
\begin{pmatrix}
0 & B \\
-B^\dagger & 0
\end{pmatrix} \label{eq:U0}
\ee
where $S_0$ is a real diagonal matrix with $[S_0]_{j}=[S_0]_{j+m+1}\,(j=1,2,\ldots,m)$ and $B$ is of size $(m+1)\times m$ and complex in general. Only the diagonal of $B$ is nonzero, i.e., $B_{ij} = \delta_{ij}b_j\,(j=1,2,\ldots,m)$, and it can be easily shown that $U_0$ of this form is indeed unitary, with $|b_j|^2+[S_0]_j^2=1$ and $[S_0]_{m+1}=1$. When the couplings are taken to be real to satisfy $T$ symmetry, $B$ becomes imaginary and turns $U_0$ into a symmetric matrix.

Similarly, we find 
\be 
U_{2} = S_{2} + 
\begin{pmatrix}
0 & A \\
-A^\dagger & 0
\end{pmatrix}, \label{eq:U2}
\ee
where $S_{2}$ is a real diagonal matrix with $[S_2]_{j+1}=[S_2]_{j+m+1}\,(j=1,2,\ldots,m)$ and $A$ is of size $(m+1)\times m$ and complex in general. Furthermore, only the first lower diagonal of $A$ is nonzero, i.e., $A_{ij} = \delta_{i-1,j}a_j\,(j=1,2,\ldots,m)$, and we find $|a_j|^2+[S_{2}]_{j+1}^2=1$ and $[S_{2}]_{1}=1$. 

In addition, $T$ symmetry determines that $U_2^*U_{-2}=1$ [see Eq.~(\ref{eq:Tsymm})], and the unitary nature of $U_2$ means $U_2^\dagger U_2=1$. Therefore, we find 
\be 
U_{-2} = U_{2}^T = 
S_{2} + 
\begin{pmatrix}
0 & -A^* \\
A^T & 0
\end{pmatrix}.\label{eq:U-2}
\ee
When we take step 2 and $-2$ to be the same (i.e., imposing $T$ symmetry in each of the two steps), we then find that $A$ is imaginary (and $U_2$, $U_{-2}$ symmetric). 

The general forms of $U_{2}$ and $U_{-2}$ indicate that they commute with
\be 
\Gamma = \text{Diag}[1,f_1,f_2,\ldots,f_m,f_1,f_2,\ldots,f_m] \label{eq:Gamma}
\ee
and its inverse, where $f_j$'s $(j=1,2,\ldots,m)$ are arbitrary non-zero complex numbers. Therefore, if we perform a similar transformation of $U$ using $\Gamma$ of this form, we find
\be 
\Gamma U \Gamma^{-1} = U_2 (\Gamma U_1) U_0 (\Gamma U_1)^{-1} U_{-2}.\label{eq:goal}
\ee   
Clearly, if $U_1$ is proportional to the inverse of $\Gamma$, then we find
\be 
\Gamma U \Gamma^{-1} = U_q, \label{eq:STrans}
\ee   
which leads to forbidden $T$-symmetry breaking of our time-dependent non-Hermitian lattice. 

\begin{figure}[t]
\centering
\includegraphics[width=\linewidth]{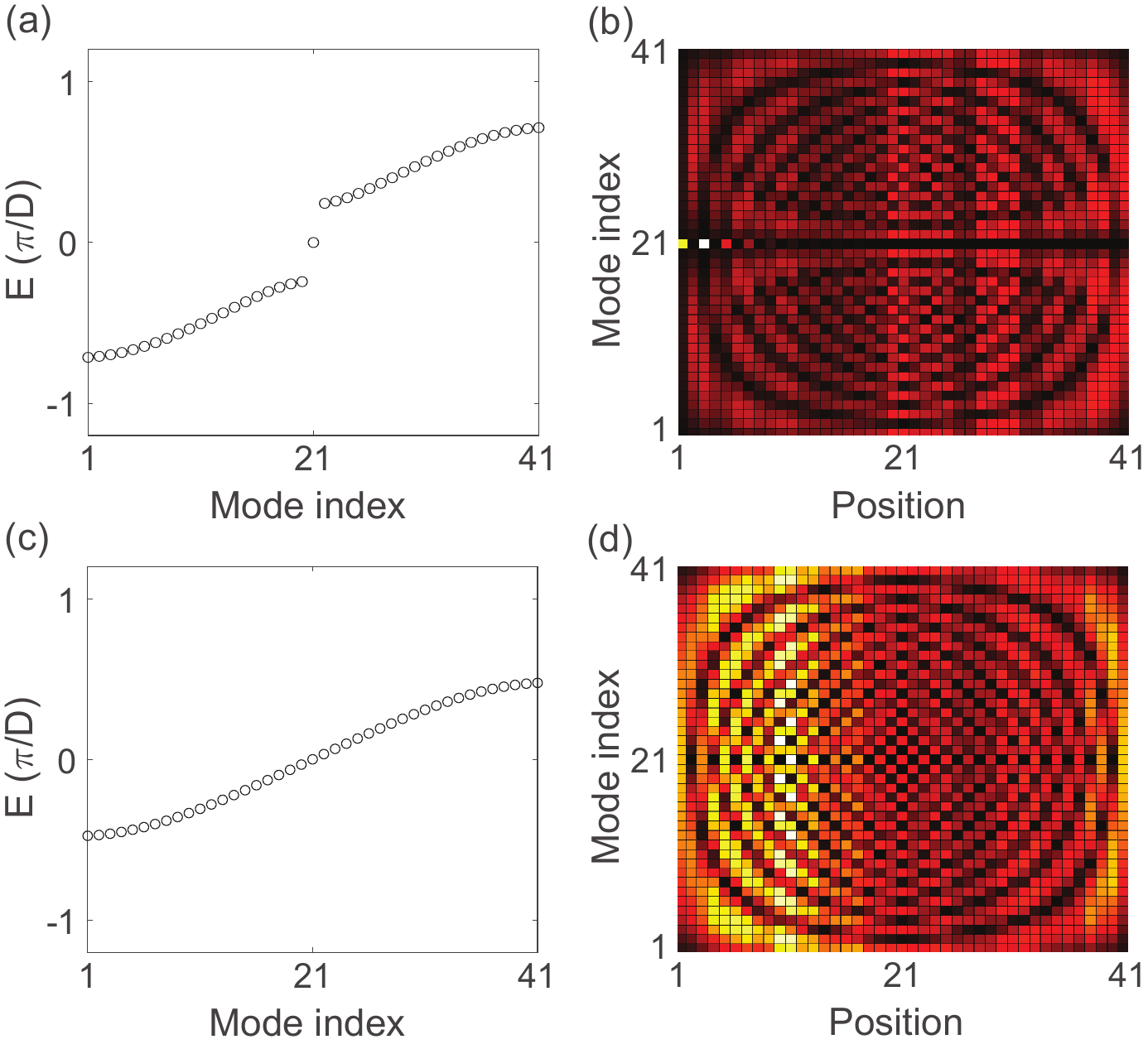}
\caption{\textbf{Forbidden $T$-symmetry breaking without the non-Hermitian skin effect}. (a,b) Scenario 1 with random but pairwise identical gain and loss in step $-1$ (and step 1) specified by Eq.~(\ref{eq:special}). The NN coupling is uniform everywhere in step 0 and given by $C$. Other parameters are the same as in Figs.~\ref{Fig:3}(a,b). (c,d) Scenario 2 with random and uncorrelated gain and loss everywhere in step $-1$ (and step $1$). The NN coupling remains staggered and equal to $C$ in step 0. Other parameters are the same as in (a,b).}
\label{Fig:4}
\end{figure}

This scenario (``scenario 1''), however, requires pairwise identical gain or loss 
\be 
r_{j+1} = g_j\,(j=1,2,\ldots,m) \label{eq:special}
\ee 
in \textit{each} non-Hermitian step, i.e., between sites 2 and 3, 4 and 5, etc. in space, even though the value of each pair is arbitrary (together with $r_1$). Therefore, this is not the case in the example shown in Fig.~\ref{Fig:3}. Nevertheless, this scenario of forbidden $T$-symmetry breaking imposes no restriction on the unitary step 0 as long as it has $T$ symmetry. For example, we can make the NN coupling everywhere and uniform in step 0, which does not destroy the forbidden $T$-symmetry breaking [see Figs.~\ref{Fig:4}(a,b)]. 

This case also indicates that the forbidden $T$-symmetry breaking is not tied to the non-Hermitian skin effect shown in Fig.~\ref{Fig:3}. Note that the staggered couplings in steps $-2$ and $2$ and the uniform coupling in step $0$ now leads to weaker effective coupling between sites 1 and 2, 3 and 4, etc. compared to those between 2 and 3, 4 and 5, etc. Therefore, we observe a dimerization effect just as in the Su–Schrieffer–Heeger chain, leading to a zero mode (mode 21) localized on the left side of the chain in Fig.~\ref{Fig:4}(b). This edge mode is not present when the coupling in step $0$ is still staggered [see Figs.~\ref{Fig:3} and \ref{Fig:4}(c,d)].

Now we focus on the more general case where $r_i$'s and $g_i$'s in $U_1$ are uncorrelated (``scenario 2''). Utilizing the the structure of $U_0$ given by Eq.~(\ref{eq:U0}), we identify $\Gamma$ with 
\be
f_j = \prod_{i=1}^j \frac{r_i}{g_i} \label{eq:f}
\ee
that leaves $U_0$ unchanged after the similar transformation in Eq.~(\ref{eq:goal}): 
\be 
(\Gamma U_1) U_0 (\Gamma U_1)^{-1} = U_0.\label{eq:Utrans}
\ee
In other words, this $\Gamma$ guarantees Eq.~(\ref{eq:STrans}) and explains the forbidden $T$-symmetry breaking exemplified by Fig.~\ref{Fig:3}. This observation is also confirmed by inspecting the structure of $U$ itself (see Appendix A).

Using the Taylor expansion of $U=e^{-iH_dD}$, it is easy to see that this $\Gamma$ also performs a similar transformation of $H_d$, i.e.,
\be
\Gamma H_d \Gamma^{-1} = H_q,
\ee
where $H_q$ is Hermitian and defined by Eq.~(\ref{eq:Hermitian}). This is the generalization of the imaginary gauge transformation found in Ref.~\cite{FNHSE} in the weak gain and loss approximation.

\begin{figure}[b]
\centering
\includegraphics[width=\linewidth]{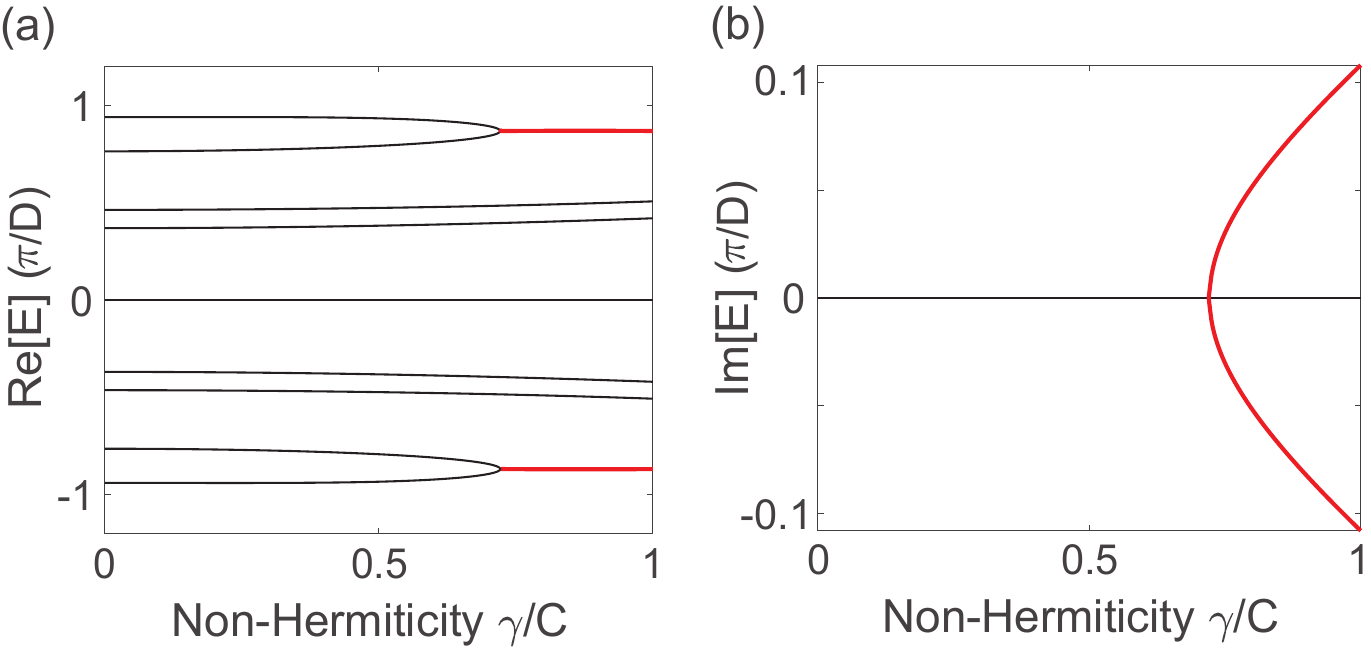}
\caption{\textbf{$T$-symmetry breaking due to non-staggered couplings}. (a,b) Real and imaginary parts of the quasi-energies. The system is the similar to that in Figs.~\ref{Fig:4}(a,b) but with 9 sites and $D=9/C$, and without the constraint given by Eq.~(\ref{eq:special}).}
\label{Fig:5}
\end{figure}

In this derivation of the general case, we have first used the condition that there is no coupling in each non-Hermitian step. 
Assuming all modes $\psi_\mu$ of the Hermitian $H_q$ are bulk states, the similar transformation $\Gamma$ changes them into $\Gamma^{-1}\phi_\mu$, i.e., superposing a spatial profile given by the diagonal elements of $\Gamma^{-1}$. Therefore, the random gain and loss presented by $\Gamma$ does not lead to non-Hermitian skin effect in general [see Figs.~\ref{Fig:4}(c,d)], and to warrant the latter, one can impose gain in one sublattice and/or loss in the other in step $-1$ (and time reverse it in step 1), with which $\{f_j\}$ becomes a monotonically increasing or decreasing series in space. This monotonic spatial profile then gives the envelope of all skin modes, which is the case shown in Fig.~\ref{Fig:3} despite the randomness of gain (loss) in step $-1$ (step 1).

The second condition we have used is that the couplings must be staggered in all unitary steps, which differs from scenario 1 given by Eq.~(\ref{eq:special}). This condition gives rise to the forms of $U_{2}$, $U_{-2}$, and $U_0$ we have used, and when it is violated, we again observe $T$-symmetry breaking as the non-Hermiticity $\gamma$ increases (see Fig.~\ref{Fig:5}). Scenarios 1 and 2 overlap when the coupling in step 0 remains staggered and the gain and loss conform with Eq.~(\ref{eq:special}), with which we find that $f_j$ given by Eq.~(\ref{eq:f}) becomes
\be 
f_j = \frac{r_1}{g_j}
\ee 
and that $\Gamma U_1$ is the identity matrix multiplied by $r_1$, again leading to Eq.~(\ref{eq:STrans}). 

Below we discuss a few generalizations of the models we have considered so far. We have taken the NN couplings in the unitary steps to be real and all equal to $C$, but they can be complex and random too, e.g., $H_2=H_2^\dagger\in\mathbb{C}$ (and $H_{-2} = H_{2}^*$), which maintain the structures of $U_2$ and $U_{-2}$ given by Eqs.~(\ref{eq:U2}) and (\ref{eq:U-2}). Therefore, we still observe the forbidden $T$-symmetry breaking (not shown). Similarly, we can also split step 0 into to two time-reversal steps given by $H_{0+}$ and $H_{0-}$, again featuring complex and random NN couplings (i.e., $H_{0+} = H_{0+}^\dagger= H_{0-}^*\in\mathbb{C}$). They do not destroy the structure of $U_0$ given by Eq.~(\ref{eq:U0}) neither, except that $S_0$ is now complex. Therefore,  the forbidden $T$-symmetry breaking also persists.

\begin{figure}[t]
\centering
\includegraphics[width=\linewidth]{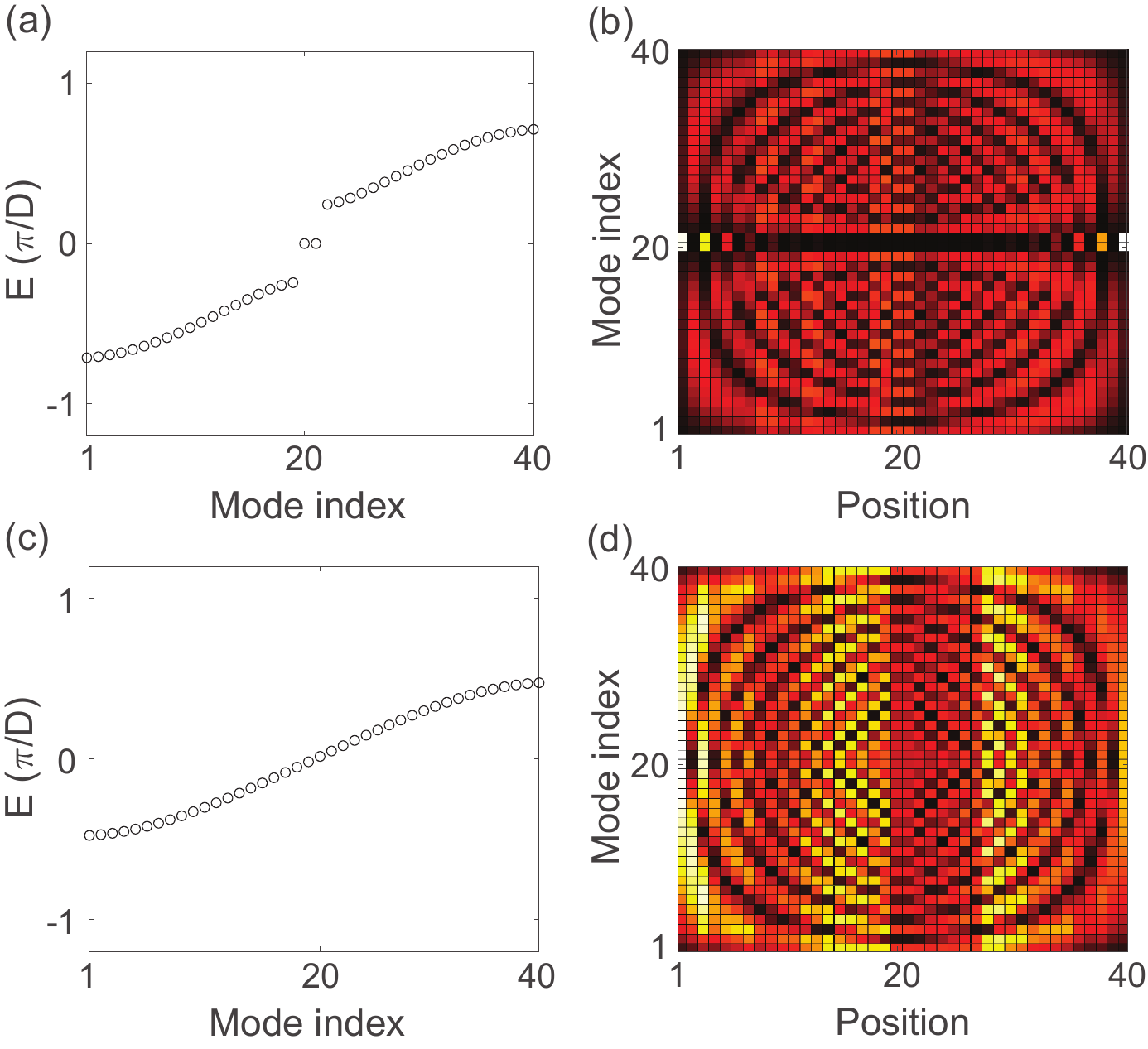}
\caption{\textbf{Forbidden $T$-symmetry breaking with an even number (40) of sites}. In (a,b) both $r_1$ and $g_{20}$ are random, while the rest are random but pairwise identical according to Eq.~(\ref{eq:special}). Other configurations are the same as in Fig.~\ref{Fig:4} for all panels. }
\label{Fig:6}
\end{figure}

We have also taken the number of lattice sites to be odd. The same forbidden $T$-symmetry breaking, nevertheless, is observed when there are an even number ($2m$) of lattice sites, e.g., by removing the last site on the right. In this latter case, scenario 1 mentioned before allows $g_n$ as a free parameter besides $r_1$, and Eq.~(\ref{eq:special}) is required for $j=1,2,\ldots,m-1$. Due to the free-standing site at each end of the chain in steps $-2$ and $2$ now, we have two quasi-zero modes localized at both ends (weakly coupled by the bulk) when the NN coupling is uniform in step 0 [modes 20 and 21 in Fig.~\ref{Fig:6}(a,b)]. 
The more general scenario 2 [see Fig.~\ref{Fig:6}(c,d)] is now enabled by a similar transformation between $U$ and $U_q$ using 
\be 
\Gamma = \text{Diag}[1,f_1,f_2,\ldots,f_{m-1},f_1,f_2,\ldots,f_{m-1},f_m]\label{eq:Gamma_even}
\ee 
where $f_j$ is still given by Eq.~(\ref{eq:f}).

\subsection{$PT$-symmetry breaking}

\begin{figure}[b]
\centering
\includegraphics[width=0.7\linewidth]{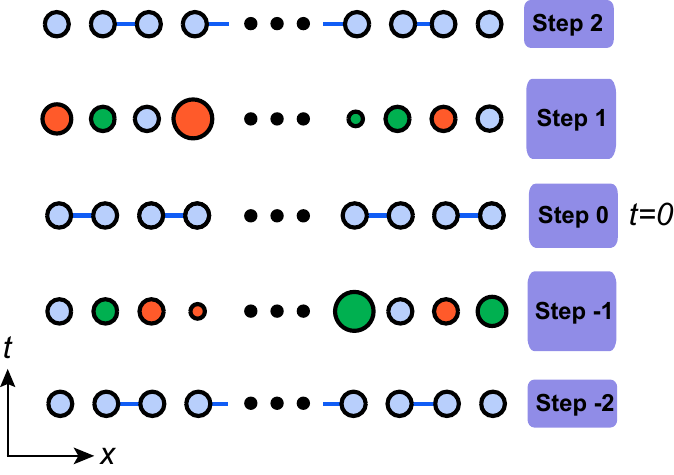}
\caption{\textbf{Example of a time-dependent non-Hermitian chain with $PT$ symmetry}. Legends are the same as in Fig.~\ref{Fig:1}.}
\label{Fig:7}
\end{figure}

The temporal modulation of our system can be easily modified to display $PT$ symmetry. For example, with an even number of sites, each unitary step can be made $PT$-symmetric by symmetrizing the staggered couplings, e.g., $C_{1,2}=C_{2m-1,2m}$ in step 0 shown in Fig.~\ref{Fig:7}, where $C_{i,j}\in\mathbb{R}$ is the coupling between sites $i$ and $j$. $PT$ symmetry of the form 
\be
PT H(x,t) = H^*(-x,-t) PT = H(x,t) PT \label{eq:PT} 
\ee 
is then achieved for the whole system with $H_2 = H_{-2}\in\mathbb{R}$ and $H_{-1}(x) = H_1^*(-x)\in\mathbb{C}$, i.e., flipping the lattice in step $1$ of our original $T$-symmetric system. We then find
\be 
UPU^* = \mathcal{T} e^{-i\int_{-D/2}^{D/2} H(x,t) dt} e^{i\int_{-D/2}^{D/2} H^*(-x,-t) dt} = P, \nonumber
\ee 
similar to how we proved Eq.~(\ref{eq:Tsymm}). Now if $\psi_\mu$ is an eigenstate of $U$, i.e., $U^*\psi_\mu^* = \lambda_\mu^*\psi_\mu^*$, then we find $UPU^*\psi_\mu^* = \lambda_\mu^* U (P\psi_\mu^*) = (P\psi_\mu^*)$, or equivalently,
\be
 U(P\psi^*_\mu) = \frac{1}{\lambda^*_\mu}(P\psi^*_\mu).
\ee 
This relation indicates that $P\psi^*_\mu$ is also an eigenstate of $U$, and there are two possible phases of $\psi_\mu$, similar to the case with $T$ symmetry. In the $PT$-symmetric phase, we have
\be 
PT\psi_\mu \equiv P\psi^*_\mu = \psi_\mu, \quad |\lambda_\mu|=1, 
\ee 
and in the $PT$-broken phase, we have instead 
\be 
PT\psi_\mu = \psi_\nu, \quad \lambda_\mu^*=\lambda_\nu^{-1}
\ee 
with $\lambda_\mu,\lambda_\nu\in\mathbb{C}\,(\mu\neq\nu)$. The corresponding eigenvalues of the effective Hamiltonian $H_d$ are again real in the $PT$-symmetric phase and form complex conjugate pairs in the $PT$-broken phase, exactly the same as a time-independent Hamiltonian.

To separate the effects of $T$ and $PT$ symmetries, we note that this system also has $T$ symmetry only when it has parity also in the non-Hermitian steps, i.e., $H_1(x) = H_1(-x)$ and $H_{-1}(x) = H_{-1}(-x)$, which then leads to $H_{-1}(x) = H_1^*(x)$. 

\begin{figure}[t]
\centering
\includegraphics[width=\linewidth]{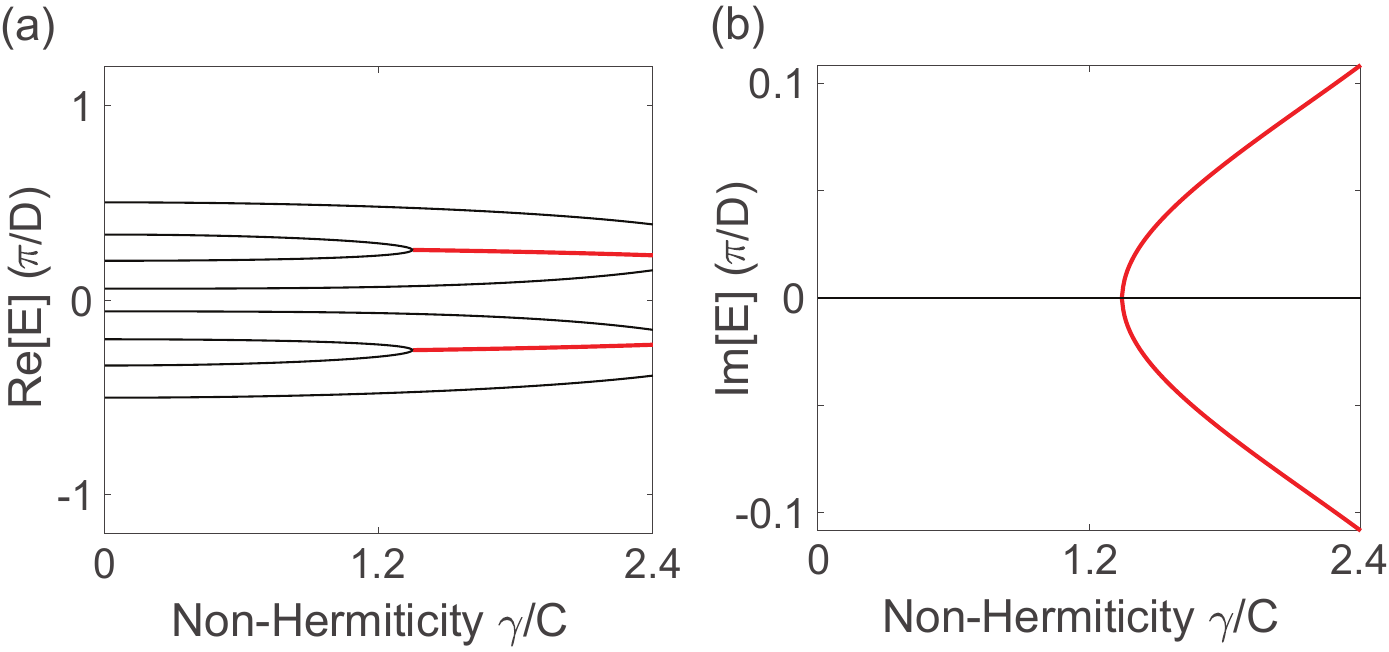}
\caption{\textbf{Absence of forbidden $PT$-symmetry breaking in a temporarily modulated lattice}. (a,b) Real and imaginary parts of the quasi-energies for a system shown in Fig.~\ref{Fig:7} with 8 sites, random staggered NN couplings (respecting $PT$ symmetry) in the unitary steps, and random gain and loss in $H_{-1}(x)=H_1^*(-x)$.}
\label{Fig:8}
\end{figure}

When $T$ symmetry is absent, we do not observe a forbidden $PT$-symmetry breaking in this system (Fig.~\ref{Fig:8}), for either scenario 1 and 2 mentioned previously.  We note that even if a forbidden $PT$-symmetry breaking existed, it could not take place in the fashion similar to scenario 1 of the $T$-symmetric case. This scenario would require $\Gamma$ to be the inverse of both $U_1$ and $U_{-1}^{-1}$, or in other words, $U_1U_{-1}=1$, which indicates $T$ symmetry of the two non-Hermitian steps (and of the entire time-dependent system). 

When the gain and loss are uncorrelated in each non-Hermitian step, i.e., in scenario 2 of the $T$-symmetric case, the absence of a forbidden $PT$-symmetry breaking can be understood in the following way. With the $PT$-symmetry relation $H_{-1}(x) = H_1^*(-x)$, and consequently, $U_{-1} = PU_1^{-1}P$, the evolution matrix $U$ of this $PT$-symmetric system has the same eigenvalues as 
\be 
U' = U'_{2}U_{1} U'_0 U_{1}^{-1} U'_{-2}, \label{eq:UPT}
\ee 
where $U'_2 = U'_{-2}$ and 
\be 
U'_0\equiv U_0P, \quad (U'_2)^2\equiv P(U_2)^2.\nonumber
\ee 
$P$ here takes the form of an anti-diagonal matrix with 1's, and $U'_0$, $U'_2$ still represent two individually $T$-symmetric steps (see an example in Appendix B). Therefore, the temporal evolution of this $PT$-symmetric system, after duration $D$, is identical to a $T$-symmetric system with $U'_0$ and $U'_{2}$ being the evolution matrices in steps 0 and $-2,2$, respectively. 

$U'_0$ and $U'_2$, however, have very different forms from $U_0$ and $U_2$ given by Eqs.~(\ref{eq:U0}) and (\ref{eq:U2}), which indicates that the corresponding Hamiltonian does not feature staggered NN couplings. Therefore, we expect that the effective Hamiltonian of this equivalent system with $T$ symmetry to acquire complex conjugate quasi-energies as the non-Hermiticity increases, and in turn, that the corresponding $PT$-symmetry system cannot remain in its $PT$-symmetric phase, as Fig.~\ref{Fig:8} depicts. 


It should be noted that we can, in principle, construct a system that displays forbidden $PT$-symmetry breaking, for a specific duration or period of temporal modulation, using a transformation of a $T$-symmetric system displaying forbidden $T$-symmetry breaking (see Appendix B). However, the resulting unitary steps would include detunings and long-range couplings that are difficult to achieve alongside the temporal modulation, for example, using the coupled waveguide system \cite{FNHSE}. More importantly, this forbidden $PT$-symmetry breaking with increasing gain and loss is lost once we change the duration or period of temporal modulation. Therefore, we do not discuss this possibility of forbidden $PT$-symmetry breaking in detail here.    

\section{Conclusion and Discussion} 

In summary, we showed that the typical scenario of non-Hermitian symmetry breaking in a finite Hilbert space is forbidden in a class of time-dependent systems with gain and loss and $T$ symmetry: all the quasi-energies of the effective Hamiltonian remain real as we increase the strength of gain and loss. This behavior is independent of the spatial distribution of gain and loss, which are implemented in steps sandwiched by unitary evolutions of the system. Furthermore, this behavior is also independent of the duration or period of the temporal modulation. 

Our findings show that expectations from time-independent non-Hermitian systems may not hold in time-dependent systems. In particular, one needs to reexamine the belief that non-Hermitian symmetry breaking in a finite Hilbert space takes place when gain and loss become comparable to certain coupling strength, which is true, for example, in the cases of $PT$-dimers and chains \cite{ge_effective_2024}, edge states in the non-Hermitian Su–Schrieffer–Heeger model \cite{song_breakup_2019}, and continuous photonic crystals \cite{makris_beam_2008}. 

This belief is based on the complementary roles of gain/loss and couplings in the formation of EPs: the former provides a diagonal term in the Hamiltonian while the latter an off-diagonal one, which cancel each other at an EP in the calculation of the eigenvalues of the Hamiltonian. This physical picture largely survives in previously studied time-dependent non-Hermitian systems, where both effective gain/loss and couplings are modified by the temporal modulation of the system \cite{li_observation_2019}. In our system, however, such temporal modifications lead back to a Hamiltonian with real eigenvalues, and the increase of non-Hermiticity does not change the characteristic polynomial of this Hamiltonian, leading to the forbidden $T$-symmetry breaking. Our findings create new possibilities and opportunities for future investigations in time-dependent non-Hermitian systems. For example, it remains to be seen whether such a forbidden non-Hermitian symmetry breaking could also take place in scattering systems, independent of the material non-Hermiticity and temporal modulation period. 

This work is supported by the National Science Foundation (NSF) under Grant No. ECCS-2602409 and a PSC-CUNY research award from the City University of New York under Grant No. GR-00018039. 

\appendix

\section{Full evolution matrix}

Below we derive the explicit form of the full evolution $U$ with an odd number of sites that exhibits the general forbidden $T$-symmetry breaking. With the two conditions specified in the main text and the individual evolution matrices given by Eqs.~(\ref{eq:U1})-(\ref{eq:U-2}), we find
\be
U_1U_0U_{-1} = S_0 + 
\begin{pmatrix}
0 & B' \\
B'' & 0
\end{pmatrix} 
\ee
where $B'_{ij} = \delta_{ij} b_jr_j/g_j$ and $B''_{ij} = -\delta_{ij} b_j^*g_j/r_j$. This result enables us to write the full evolution matrix as
\be
U = U_s + U_g(\gamma),  
\ee
where $U_s$ does not depend on gain and loss and appear in the unitary evolution 
\be 
U_q=U_{2}U_0U_{-2} = U_s + U_g(0) 
\ee 
as well, i.e., where there is no gain or loss in steps $-1$ and $1$. $U_s$ is given by
\begin{align} 
U_s = S_0S_2^2 &
+
\begin{pmatrix}
0 & A \\
-A^\dagger & 0
\end{pmatrix}S_0S_2
+ S_0S_2
\begin{pmatrix}
0 & -A^* \\
A^T & 0
\end{pmatrix} \nonumber \\
&+
\begin{pmatrix}
0 & A \\
-A^\dagger & 0
\end{pmatrix}
S_0
\begin{pmatrix}
0 & -A^* \\
A^T & 0
\end{pmatrix}.
\end{align}
$U_g(\gamma)$ depends on the non-Hermiticity $\gamma$, and it is given by
\begin{align}
U_g(\gamma) = \,& S_2\begin{pmatrix}
0 & B' \\
B'' & 0
\end{pmatrix}S_2 -
\begin{pmatrix}
0 & AB''A^* \\
A^\dagger B'A^T
\end{pmatrix} \nonumber \\
& +
S_2
\begin{pmatrix}
AB'' + B'A^T& 0 \\
0 & -(A^\dagger B'+B''A^*)
\end{pmatrix}.  \nonumber
\end{align}
In it, the first term represents NN evolution between sites 1 and 2, 3 and 4, etc. in space, coming from $U_0$. The second term describes the NNN evolution (i.e., within each sublattice), and the last term gives the next-order evolution, between sites 2 and 5, 3 and 6, etc. in space. Denoting $U_g$ by
\be 
U_g(\gamma) = 
\begin{pmatrix}
U_{ul} & U_{ur} \\
U_{bl} & U_{br}
\end{pmatrix}
\ee  
where $U_{ul}$ is a $(m+1)\times(m+1)$ matrix and $U_{br}$ is $m\times m$ matrix, we find that the sequence $p_j\equiv r_j/g_j\,(j=1,2,\ldots,m)$ or its inverse appears in all non-zero elements in $U_g$: The first lower diagonal of $U_{ul}$ and the diagonal of $U_{bl}$ are proportional to $p_1,p_2,\ldots,p_m$. The first upper diagonal of $U_{ul}$ and the diagonal of $U_{ur}$ are proportional to $p_1^{-1},p_2^{-1},\ldots,p_m^{-1}$. The first lower diagonal of $U_{br}$ and the second lower diagonal of $U_{ur}$ are proportional to $p_2,p_3,\ldots,p_m$. The first upper diagonal of $U_{br}$ and the second upper diagonal of $U_{bl}$ are proportional to $p_2^{-1},p_3^{-1},\ldots,p_m^{-1}$. 

These property enables a similar transformation between $U_g(\gamma)$ and $U_g(0)$:
\be 
U_g(0) = \Gamma U_g(\gamma)\Gamma^{-1}.
\ee
This $\Gamma$ is the one given by Eqs.~(\ref{eq:Gamma}) and (\ref{eq:f}) in the main text. Furthermore, we note that this structure of $\Gamma$ also leaves $U_s$ unchanged, e.g.,
\begin{align}
\Gamma\begin{pmatrix}
0 & A \\
-A^\dagger & 0
\end{pmatrix}S_0S_2\Gamma^{-1}
&= 
\Gamma\begin{pmatrix}
0 & A \\
-A^\dagger & 0
\end{pmatrix}\Gamma^{-1}S_0S_2
\nonumber \\
&=
\begin{pmatrix}
0 & A \\
-A^\dagger & 0
\end{pmatrix}S_0S_2, 
\end{align}
because only the first lower diagonal of $A$ is nonzero. Therefore, we recover the general case of forbidden $T$-symmetry breaking discussed in the main text, i.e., this $\Gamma$ performs a similar transformation between $U$ and the unitary evolution $U_q$:
\be 
U_q = \Gamma U\Gamma^{-1}.
\ee

\section{Forbidden $PT$-symmetry breaking}

As mentioned in the main text, we can construct a system that displays forbidden $PT$-symmetry breaking, for a given temporal duration or period, using a transformation of a $T$-symmetric system displaying forbidden $T$-symmetry breaking, which however requires long-range couplings and detunings. 

To elucidate this observation, we first note that $U$ in Eq.~(\ref{eq:U}) of the $T$-symmetric evolution matrix has the same eigenvalues as
\be 
\tilde{U} = U_{1} (U_0P) (PU_{1}^{-1}P) (PU_{-2}U_2).
\ee 
Note again that the parity operator satisfies $P^2=1$, and we have moved $U_2$ from the leftmost position in $U$ to the rightmost position in $\tilde{U}$. We also note that our original steps 2 and $-2$ are the same, and hence $U_{-2}U_2$ is the evolution with the Hermitian Hamiltonian $H_2$ for a duration of $2D_2$. We then denote
\be 
U_0P\equiv U'_0 \equiv e^{-iH'_0D_0},\quad PU_{-2}U_2 \equiv e^{-2iH'_2D_2} = (U'_2)^2\nonumber
\ee 
and define
\be 
U' = U'_2 U_1 U'_0 U'_{-1} U'_{-2},\label{eq:U_new2}
\ee 
which has the same eigenvalues as $\tilde{U}$ and $U$. Here $U'_2=U'_{-2}$ and $U'_{-1}$ is defined by
\be 
 U'_{-1}\equiv PU_{1}^{-1}P\equiv e^{-iH'_1D_1},
\ee
and this step is the $PT$ reversal of step 1.

We now have a new time-dependent system described by $U'$ with five steps that displays real quasi-energies independent of the spatial distribution and strength of gain and loss. Both step 2 ($-2$) and step 0 (and hence the entire system) can be made $PT$-symmetric when the original $U_2,U_0$ are also $P$-symmetric, leading to forbidden $PT$-symmetry breaking as gain and loss increases. 

As we show below, these $H'_2,H'_0$, however, now include detunings and long-range couplings due to $P$ in the definitions of $U'_2,U'_0$, which are difficult to achieve alongside the temporal modulation, e.g., on the coupled waveguide system \cite{FNHSE}. Furthermore, these elements depend on the duration $D_2$ and $D_0$, which means that such forbidden $PT$-symmetry breaking is lost when we change these temporal durations.

To understand these observations, we first note that $P$ can be written as 
\be 
P = e^{-i X_j} \,(j=0,2)
\ee 
where $X_j$ is a $2n\times2n$ matrix, i.e., with the same size as $P$. Because $P^2=1$, $1$ and $-1$ are the only two eigenvalues of $P$, each with an algebraic degeneracy of $n$. For each degenerate subspace of the associated Hilbert space, one can choose an arbitrary set of orthogonal eigenstates $\{\phi^{(\mu)}_m\}\,(\mu=\pm; m=1,2,\ldots,n)$ of $P$, i.e.,
\be 
P \phi^{(\mu)}_m = \mu \phi^{(\mu)}_m,
\ee 
which are also eigenstates of $X_j$:
\be 
X_j \phi^{(\mu)}_m = i\mu \phi^{(\mu)}_m.
\ee 
For example, $\{\phi^{(\pm)}_m\}$ can be chosen as
\be 
\phi^{(\pm)}_1 = \frac{1}{\sqrt{2}}
\begin{bmatrix}
1 \\
0 \\
0 \\
\pm1
\end{bmatrix},\,
\phi^{(1)}_2 = \frac{1}{\sqrt{2}}
\begin{bmatrix}
0 \\
1 \\
\pm1 \\
0
\end{bmatrix}
\ee 
when $n=2$. By defining
\be 
Q_j = [\phi^{(+)}_1 \phi^{(+)}_2 \ldots \phi^{(+)}_n \phi^{(-)}_1 \phi^{(-)}_2 \ldots \phi^{(-)}_n]
\ee  
that satisfies $Q_j^TQ_j=1$, we can write $X_j$ as
\be 
X_j = iQ_j\,\text{Diag}[1,1\ldots,1,-1,-1,\ldots,-1]\,Q_j^T.
\ee  
Because of the freedom of choosing $Q_j$, we can choose $Q_j$ such that 
\be 
[X_j,H_j]=0\quad (j=0,2).
\ee 
As a result, we find
\be 
H'_0 = H_0 + D_0^{-1}X_0, \quad H'_2 = H_2 + (2D_2)^{-1}X_2, \label{eq:Hp}
\ee 
where the extra factor of $2$ in the denominator of $H'_2$ is due to the product of $U_{-2}$ and $U_2$ in $\tilde{U}$ given in Eq.~(\ref{eq:U_new2}).

\begin{figure}[t]
\centering
\includegraphics[width=\linewidth]{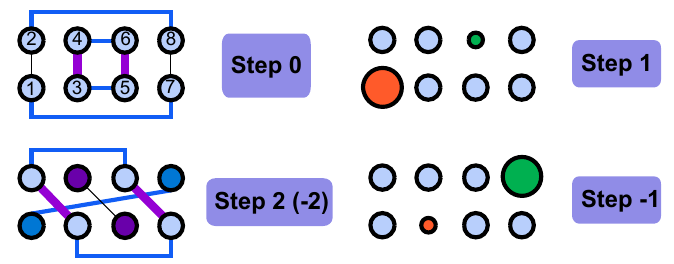}
\caption{\textbf{Example of a time-dependent non-Hermitian chain with forbidden $PT$-symmetry breaking for given temporal durations $D_0$ and $D_2$ in steps 0 and 2}. Numbers in step 0 mark the positions of the sites in the original $T$-symmetric 1D lattice, with site 1 on the left and site 8 on the right. Lines indicate real and reciprocal couplings, and darker dots in step 2 (-2) indicate detunings. Green (red) dots of different sizes represent gain (loss) of various strength in steps $-1$ and $1$, which are transformed into each other by $PT$, where $P$ takes the form of a $180^\circ$ rotation.}
\label{Fig:2D}
\end{figure}

We illustrate the resulting long-range couplings and detunings in Fig.~\ref{Fig:2D} with 8 sites, where we have rearranged the 1D system into two rows with a slightly more viable coupling scheme. In this rearrangement, the parity operator takes the form of a $180^\circ$ rotation, exchanging, for example, sites 1 and 8 as well as sites 3 and 6.

In this case, $X_0$ can be chosen as
\be 
X_0 = 2i\pi
\begin{pmatrix}
P_n & 1 \\
1 & P_n
\end{pmatrix},
\ee
where $P_n$ is the parity operator for a 1D chain of $n=4$ sites, i.e., an anti-diagonal matrix with 4 $1$'s. Therefore, we find that the central four sites form a ring in step 2 ($-2$), and so does the outer four sites. The reciprocal couplings within each row are the same and given by $2\pi/D_0$, while the rest are paired by $PT$ but random, i.e., $C_{3,4}=C_{5,6}$ that differ from $C_{1,2}=C_{5,6}$.

\begin{figure}[t]
\centering
\includegraphics[width=\linewidth]{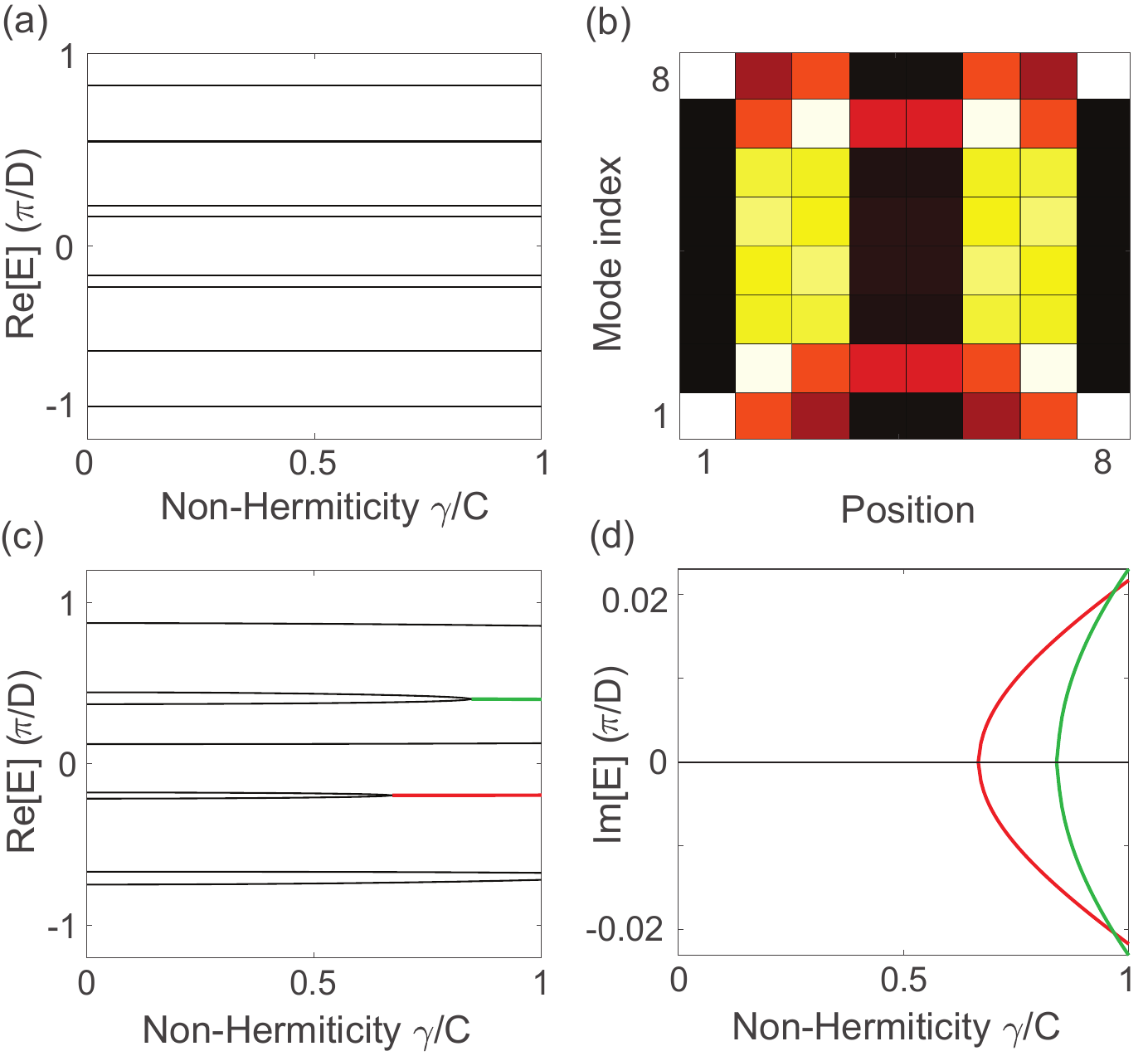}
\caption{\textbf{Forbidden $PT$-symmetry breaking and its absence}. (a) Real quasi-energies for the system shown in Fig.~\ref{Fig:2D}, with random gain and loss in step 1 (and -1). $D=9/C$ and $D_i=D/4\,(i=-1,0,1),\, D/8\, (i=-2,2)$. (b) Spatial profile $|\psi_\mu|$ for all modes at $\gamma=1$. (c,d) Complex quasi-energies as a function of the non-Hermiticity. The system is the same as (a) but with $D=6$.}
\label{Fig:PT}
\end{figure}

$X_2$ is slightly more complicated, and the resulting NN couplings between the two rows in step 2 (-2) become slated. They remain random except for $C_{1,8}=\pi/D_2$, and the latter equals the couplings with each row, i.e., $C_{3,7}$ and $C_{2,6}$. Now we also have 4 detunings at sites 1, 4, 5, and 8, all real and again with amplitude $\pi/D_2$. 

We show the persisting real quasi-energies in Fig.~\ref{Fig:PT}(a) for this system when we increase the strength of gain and loss, and the spatial profiles of the corresponding eigenstates are shown in Fig.~\ref{Fig:PT}(b). The symmetric mode profiles verify that all modes are indeed in their $PT$-symmetric regime. 

Again, this forbidden $PT$-symmetry breaking persists when we change the couplings in $H'_0$ and $H'_2$ ($H'_{-2}$) that are independent of $D_0$ and $D_2$ in the original $T$-symmetric system (e.g., $C_{3,4}=C_{5,6}$ in step 0), and it is also independent of the spatial distribution of gain and loss in steps 1 and $-1$. However, it is lifted, as mentioned before, when we use a different set of values for $D_0$ and $D_2$ in the now $PT$-symmetric system (e.g., reduced by a third; see Figs.~\ref{Fig:PT}(c,d)] \textit{while} keeping the Hamiltonian in each step fixed. 

\newpage 
\bibliography{references2}

\begin{thebibliography}{22}
\bibitem[{bib()}]{bibnote:1}
\bibinfo{note}{In principle, we have $C(H_dD)^*C = -H_dD + 2m\pi$, where $m$ is
  an arbitrary integer. In order for this relation to hold as we vary the
  temporal duration $D$, we know that $m$ must be zero, which gives
  $\{H_d,CK\}=0$.}
\end{thebibliography}

\end{document}